\documentclass[preprint,12pt,sort&compress]{elsarticle}

\usepackage{amssymb}
\usepackage{amsthm}
\usepackage{color}
\DeclareRobustCommand*\textmu{\leavevmode{\usefont{U}{psy}{m}{n} \symbol{"6D}}}

\usepackage{lineno}

\journal{arXiv}

\begin{document}

\begin{frontmatter}



\title{A new method to correct deformations in emulsion using a precise photomask}


\author[toho]{M.~Kimura\corref{cor1}\fnref{fn1}}
\ead{mitsuhiro.kimura@lhep.unibe.ch}
\author[toho]{H.~Ishida}
\author[toho]{H.~Shibuya}
\author[toho]{S.~Ogawa}
\author[toho]{T.~Matsuo}
\author[toho]{C.~Fukushima}
\author[toho]{G.~Takahashi}
\author[chiba]{K.~Kuge}
\author[utsunomiya]{Y.~Sato}
\author[utsunomiya]{I.~Tezuka}
\author[nihon]{S.~Mikado}
\address[toho]{Department of Physics, Toho University, Miyama, Funabashi, 274-8510, Japan}
\address[chiba]{Chiba University, Chiba, 263-8522, Japan}
\address[utsunomiya]{Utsunomiya University, Utsunomiya, 321-8505, Japan}
\address[nihon]{Nihon University, Narashino, 275-8576, Japan}
\cortext[cor1]{Corresponding author}
\fntext[fn1]{Now at Albert Einstein Center for Fundamental Physics, Laboratory for High Energy Physics (LHEP), University of Bern, CH-3012 Bern, Switzerland}
\begin{abstract}
A new method to correct the emulsion deformation, mainly produced in the development process,
is developed to recover the high accuracy of nuclear emulsion as a tracking device. 
The method is based on a precise photomask and a careful treatment of the emulsion films.
A position measurement accuracy of 0.6 \textmu m is obtained over an area of 5 cm $\times$ 7 cm.
The method allows to measure positions of track segments with submicron accuracy
in an ECC brick with as few as 10 reference tracks for alignment.
Such a performance can be important for hybrid emulsion experiments
at underground laboratories where only a small number of reference tracks for alignment are available.
\end{abstract}


\end{frontmatter}

\linenumbers

\section{Introduction}
Nuclear emulsions have been used as a particle detector for their excellent submicron
spatial resolution, e.g. in experiments studying short-lived particles
such as charm and bottom hadrons~\cite{Niu,E531,WA75,E653} or
double-$\Lambda$ hypernuclei~\cite{E176,E373}.
Some recent neutrino experiments also make use of
the nuclear emulsions in a hybrid apparatus~\cite{DONUT,CHORUS,OPERA}. In such experiments, both production and decay vertices
of short-lived particles are reconstructed from the measurements of the parent and daughter
particle tracks in nuclear emulsion films. However the practical application of its excellent
spatial resolution is limited to an area of a few mm$^2$ in the vicinity of the interaction
and decay vertices because of deformation of emulsion films due to various reasons
such as the thermal expansion and distortion of the emulsion layers. 
This deformation deteriorates the track position measurement accuracy over a larger area.
For example, the momentum of a charged particle can be estimated by measuring precisely
its multiple Coulomb scattering(MCS) in an Emulsion Could Chamber(ECC) brick
consisting of emulsion films interleaved with thin metal plates.
For a track with a large angle with respect to the perpendicular to the emulsion films, the measurement of its MCS requires
precise track position measurement over a large area, which is affected by the deformation
of emulsion films~\footnote{There are two methods for the MCS measurement in an ECC brick;
the ``angular" method~\cite{Angular,Angular2} and the ``coordinate" method~\cite{Coordinate}. In this paper, 
the latter is considered because the application of the former method is limited
to low momentum particles.}. 
Usually, the effect of the emulsion deformation is corrected by
using high momentum reference tracks with a similar angle 
passing near the track of interest.
However, this method cannot be applied in the case, such as in long baseline neutrino
oscillation experiments in an underground laboratory, where few references are available.
Even though cosmic ray for alignment of ECC brick are irradiated, 
the angle of their tracks are not necessarily the same as interesting track.

In this paper, we present a new method to correct the deterioration in the track 
measurement accuracy due to the deformation and 
report on the results of a muon beam exposure which was performed to demonstrate 
the performance of the method by estimating the momenta of beam muons from their MCS measurements.
The method will allow the nuclear emulsion to be used as a tracking device
in a wide range of applications such as the emulsion spectrometer technique~\cite{TES,ISS},
an emulsion detector placed in a magnetic field, 
proposed for future neutrino oscillation experiments.

\section{Concept to remove deformations in emulsion}
The major part of the deformations in emulsion originates from the
development process. Since the reference marks printed before the development 
retain their initial positions, we can know the deformations in the emulsion 
film by comparing measured positions of the reference marks with their 
designed values, then we can subtract them. By this technique, the whole 
area of an emulsion film will become a position detector with one micron 
accuracy.

We must manufacture a photomask as a printing mask to transfer the reference 
marks with submicron accuracy to the whole emulsion film. Reference marks 
in past emulsion experiments were transcribed from a negative film~\cite{CHORUS2} or 
X-ray sources with 0.1 mm slits~\cite{OPERA}. These marks had position accuracy of 
10-30 \textmu m and they were used as guides to lead the microscope view to 
the interested position such as an interaction vertex. The role of our 
reference marks is completely different from that of the marks in past 
experiments.

The arrangement of the reference marks on the photomask is designed 
according to the nature of the deformations in emulsion. The deformation 
size depends on the interested position in an emulsion films: it is small 
at the central part, but it is large at the edge part. The typical 
deformation sizes are 0.3 \textmu m within the area of several hundred \textmu m$^2$ 
at the center, 1 \textmu m in a few cm$^2$ and 10 \textmu m in one hundred cm$^2$ 
respectively.

In order to validate this correction method, we have tried to determine
the momentum of beam muons using the MCS measurement, which requires the
precise position measurement over a large area in emulsion films.

\section{Setup and beam exposure}
\begin{figure}[htbp]
 \begin{center}
  \includegraphics[width=140mm]{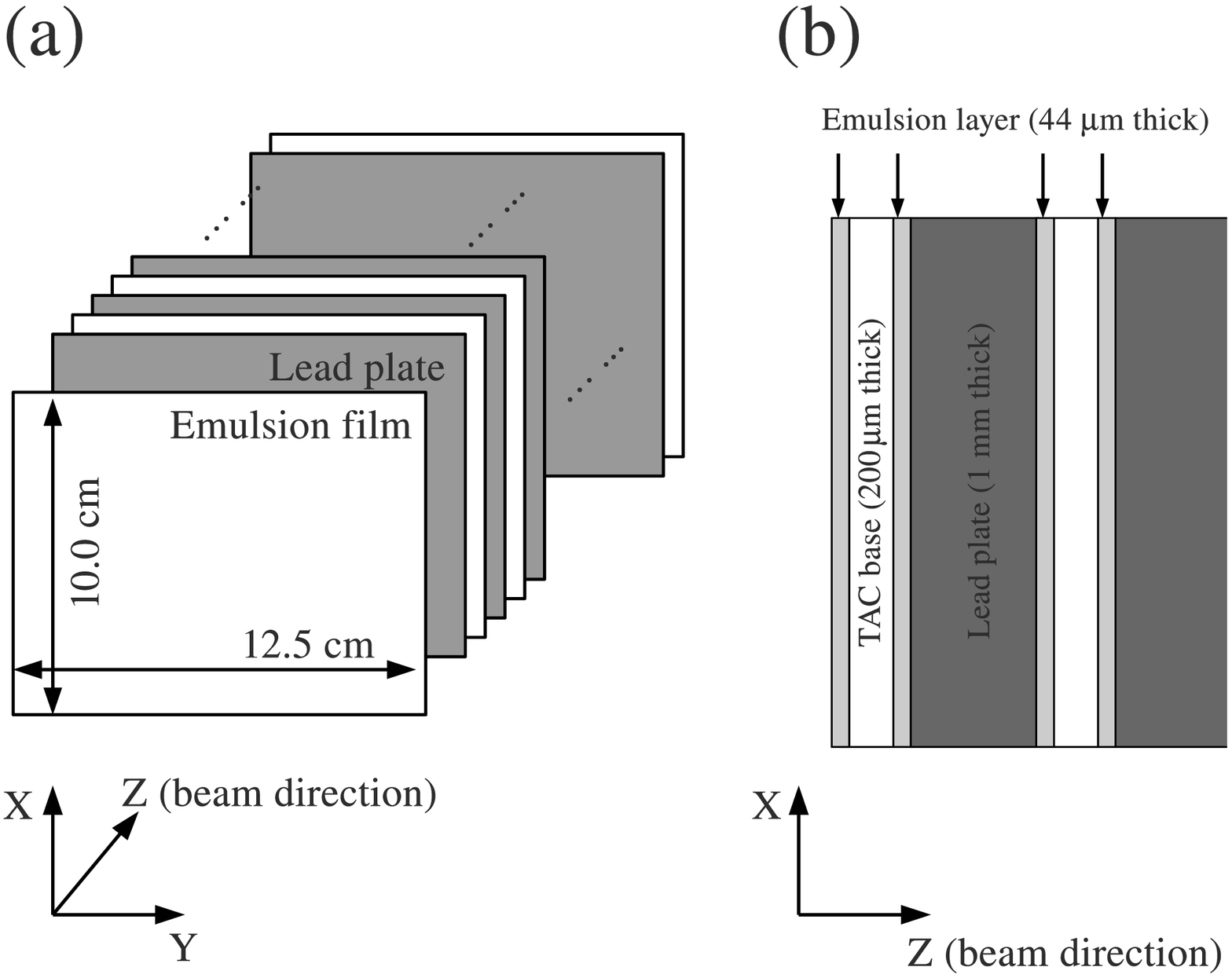}
 \end{center}
 \caption{Schematic structure of the ECC brick. (a) Overview and (b) cross view. }
 \label{fig:ecc}
\end{figure}

We carried out a muon beam exposure of an ECC brick at CERN SPS T2-H4 beam line in July 2007. 
The ECC brick was composed of 28 nuclear emulsion films interleaved with 1~mm thick lead plates
as illustrated in Fig.~\ref{fig:ecc}. The ECC brick was contained in an acrylic resin box and
enclosed in an opaque vacuum bag. Furthermore the brick was enclosed in an extruded polystyrene 
(STYROFORM\textsuperscript{\textregistered}) container for heat insulation. 
The emulsion film is the one developed 
for the OPERA experiment~\cite{Nakamura}.
It was made of a 205~\textmu m thick triacetyl cellulose (TAC) base 
with 44~\textmu m thick emulsion layers on both faces. 
The ECC brick size was 12.5~cm wide, 10.0~cm high and 3.6~cm thick 
which corresponds to 5 radiation lengths. The ECC brick was exposed to 30, 40 and 
150~GeV/$c$ muon beams and inclined horizontally to the beams with $\pm$0.3, $\pm$0.1 
and $\pm$0.2~rad respectively. A 150~GeV/$c$ muon beam was exposed perpendicularly for reference. 
The exposed beam density was controlled to be about $7\times 10^2$/cm$^2$. 
The beam density is almost uniform in the central region whose size is 3~cm $\times$ 3~cm, 
and it is lower in the surrounding regions. Full width at half maximum of beam profiles are
 around 7~cm $\times$ 7~cm.
The beam angular spread was about 1~mrad and its momentum accuracy 1\%.\par
\begin{figure}[htbp]
 \begin{center}
  \includegraphics[width=90mm]{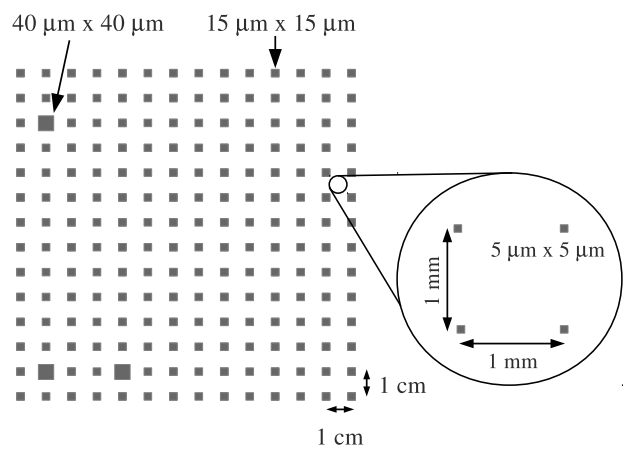}
 \end{center}
 \caption{Design pattern of grid mark on photomask. Grid marks of 5 \textmu m square shape are located with spacing of 1 mm. These marks are for the correction of track position. 15 and 400~\textmu m marks are guides to move the microscopic view center to 5~\textmu m marks.}
 \label{fig:photomaskdesign}
\end{figure}
\begin{figure}[htbp]
 \begin{center}
  \includegraphics[width=90mm]{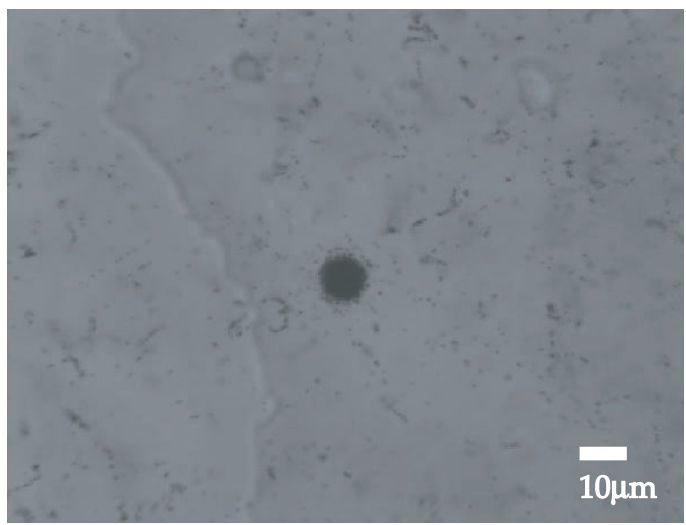}
 \end{center}
 \caption{Printed image of a 5 \textmu m grid mark on a film 
by the contact printing method.}
 \label{fig:fiducialmark}
\end{figure}
\begin{figure}[htbp]
 \begin{center}
  \includegraphics[width=140mm]{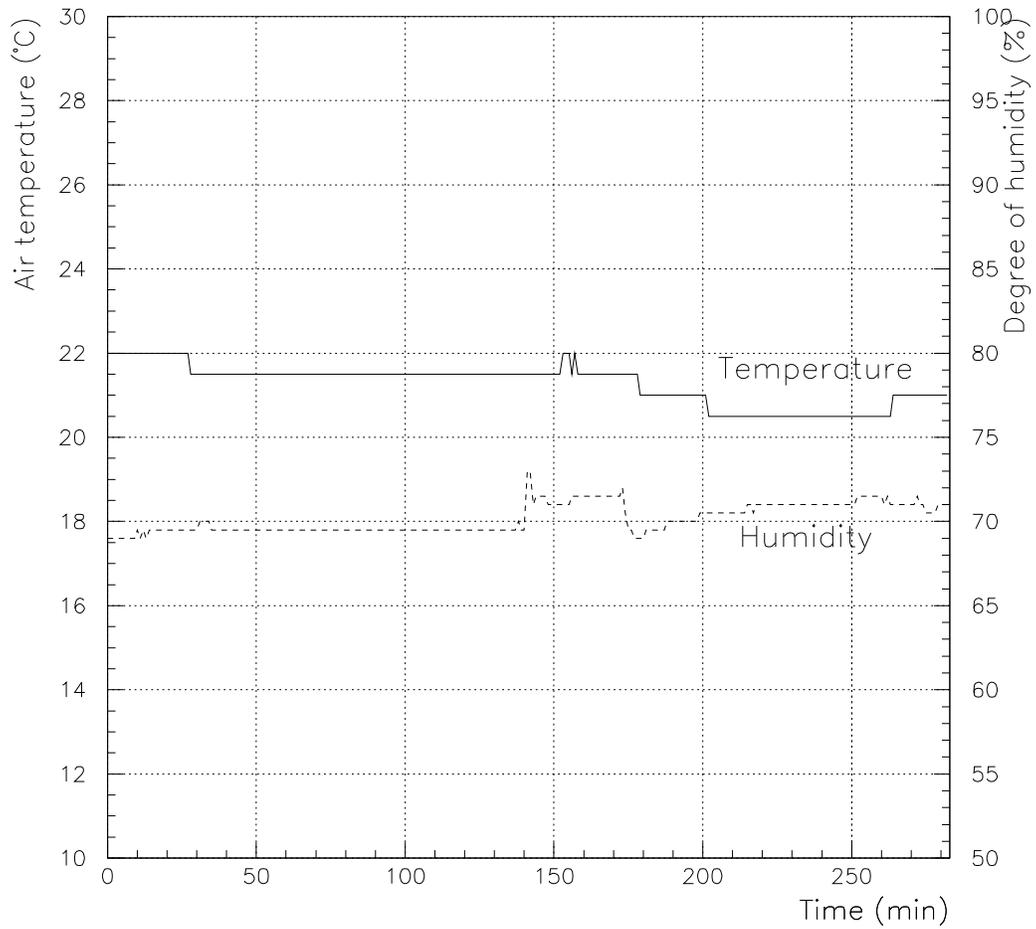}
 \end{center}
 \caption{Temperature and relative humidity of the atmosphere surrounding the ECC 
from the beginning of the beam exposure to the end of the photomask printing. 
Solid line (dashed line) indicates the temperature (humidity) transition. }
 \label{fig:condition}
\end{figure}

The emulsion handling was performed with great care. The ECC brick was dismantled
in the dark room soon after the exposure. Grid marks were printed on each emulsion film 
using a photomask~\footnote{
The photomask is a chrome coated glass lithographic template designed to optically
transfer a pattern to an emulsion layer.
The whole size of the glass substrate is 153~mm $\times$ 153~mm $\times$ 6.35~mm.
A chromium oxide layer is deposited over the glass substrate, forming a light shielding film
0.4~\textmu m thick.
Light transmittance of this layer is about 1\%.
The pattern on the chromium layer was made by using a photomechanical and etching process.
} made of synthetic quartz with a small thermal expansion
coefficient of $5.8 \times 10^{-7}$~K$^{-1}$. 
Grid marks were printed over a 127~mm $\times$ 127~mm area. 
Figure.~\ref{fig:photomaskdesign} shows the design pattern of grid marks on the photomask. 
Square shape grids, of 5~\textmu m (15~\textmu m, 400~\textmu m) side, 
were printed at intervals of 1~mm (10~mm, 100~mm). 
The accuracy of the intervals is 0.1~\textmu m. The printed grid image is 
shown in Fig.~\ref{fig:fiducialmark}. 
The printed image is not square but rather circular because the light from an electric flash
 in contact printer is not parallel, it scatters in the emulsion layer, and the contact
 between the photomask and the film is not perfect.
It took only 4.5 hours from the beginning of the exposure to the end of the grid mark printing. 
Such a fast operation minimizes the emulsion film deformation 
due to the environmental change during transportation to the processing room. 
Fig.~\ref{fig:condition} shows the temperature and 
the relative humidity variations as a function of time. 
The maximum changes in the temperature and 
humidity were 1.5$^\circ\hbox{C}$ and 4.5\% respectively. \par
\begin{figure}[htbp]
 \begin{center}
  \includegraphics[width=140mm]{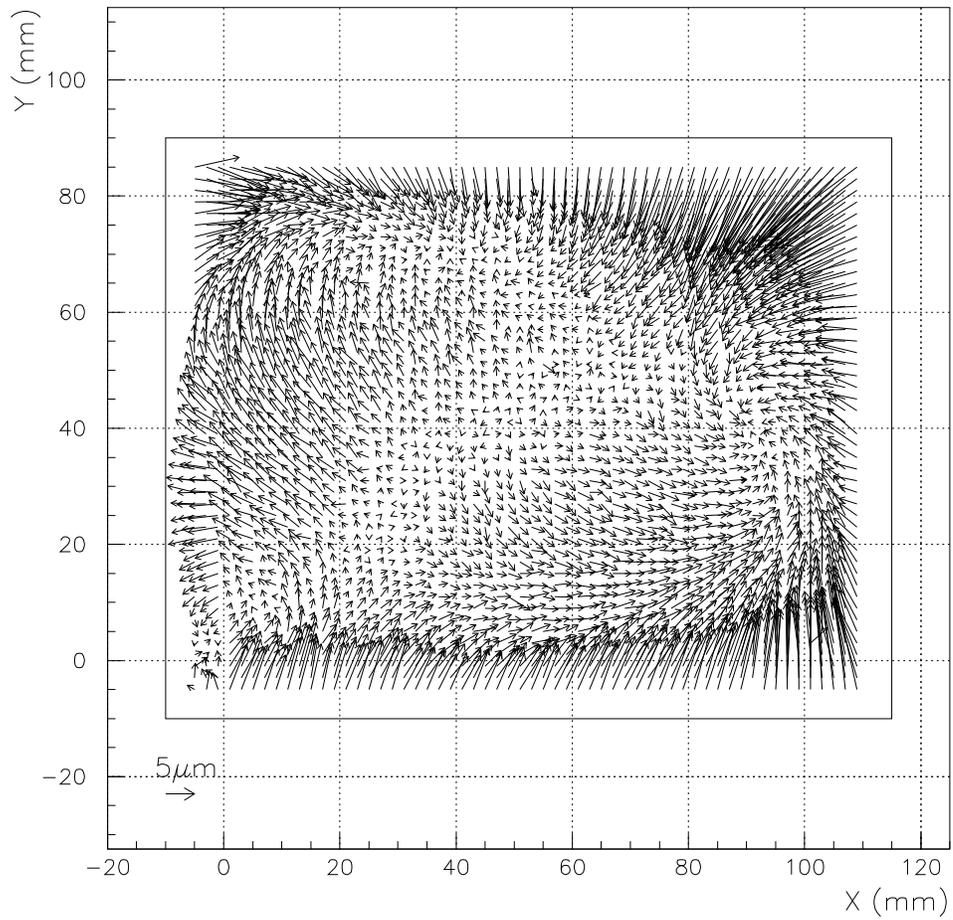}
 \end{center}
 \caption{OPERA film deformation reproduced from the grid mark measurements. 
Each vector shows the position displacement from the original coordinates of the grid mark. 
Maximum vector size is about 20 \textmu m.}
 \label{fig:deformation}
\end{figure}

For a typical film, Fig.~\ref{fig:deformation} shows the difference of the grid mark coordinates between
the original ones and the measured ones. 
Each vector shows the size and direction of the displacement, mainly caused
by the development processes. Position displacements of a few~\textmu m in the central region
and larger than 10~\textmu m near the edges are observed. 
This is an example of the deformation that the method described in this paper is aiming to correct.
\section{Emulsion measurements and analysis method}
Positions of track segments and grid marks are measured using a fully automated system called 
Ultra Track Selector (UTS)~\cite{TS,Nakano,ESS,Morishima}. 
The UTS is a microscope with a motor controlled three axes stage and an image processing unit. 
An image is taken by a CCD camera, whose pixel size corresponds to 
0.29~\textmu m$\times$0.23~\textmu m on the emulsion film. The positions of the stage 
are encoded by the LS406 of HEIDENHAIN Co.Ltd. The UTS reads out 16 tomographic images
through the 44~\textmu m thick emulsion layer, then the image processing unit 
extracts tracks with a slope smaller than 0.4~rad in the field of view. Just before 
the track segments are recorded by UTS, grid mark images are taken and automatically recognized. 
Their center positions are measured by a shape fitting algorithm.
The accuracy of grid mark center of 5~\textmu m square size is $\sigma = 0.3$~\textmu m 
in each directions. Only 5~\textmu m grid marks are used to correct the emulsion deformation 
because their positions can be measured most precisely.

An emulsion film is continuously expanding or shrinking under the influence of 
environmental temperature and humidity, even on the microscope stage. 
To get rid of this effect, we measured a well-defined reference position 
of the emulsion film at short intervals during the scanning of an area. 
We performed measurement of track segments in every 7~mm $\times$ 7~mm area at 2~cm 
intervals on the emulsion film. 
The measurement of the grid marks and the track segments in the central 1~mm $\times$ 1~mm area 
of each scanning area took about 30~minutes (10~minutes) in the central part (in the edge part) of the film. 
Position shift in the edge part by the environmental variation is greater than the central ones.
That is why the correction must be done more frequently in the edge part.
To evaluate the reproducibility of the measurements, all emulsion films were scanned 
twice in succession.
The reproducibility of the track positions is found to be 0.5~\textmu m 
at one standard deviation.\par
After the scanning, the two corresponding track segments in the emulsion layer 
on both sides were connected across the 205~\textmu m thick plastic base; 
so-called base tracks were produced. The position coordinates of each base track are 
defined as the position of the silver grain closest to the plastic base in the upstream emulsion layer. 
Base tracks are connected between consecutive emulsion films, then muon tracks are reconstructed
over the whole ECC brick. 
A connection between 2 films was accepted if the positions of the track segments agreed within 
a tolerance of about 10~\textmu m (3~$\sigma$).\par
After the connection, the positions of the base tracks of the 150~GeV/$c$ reference 
muons are determined with respect to the neighboring grid marks. 
The printed position shifts $(\Delta x, \Delta y)$ and rotation angle $\phi$
of the photomask are then calculated, for each film, as the values which
minimize the sum of the square of 2D position difference of all the pairs of base tracks 
between two consecutive films. The minimization process is done by the whole area of the film 
so as to get larger lever arm for determination of rotation angle.
 The average difference of the printed positions of the photomask are 
$\phi_\mathrm{RMS} = 3.6$~mrad 
and $(\Delta x,\Delta y)_\mathrm{RMS} = (0.30$ mm, 0.15 mm$)$.\par
\begin{figure}[htbp]
 \begin{center}
  \includegraphics[width=140mm]{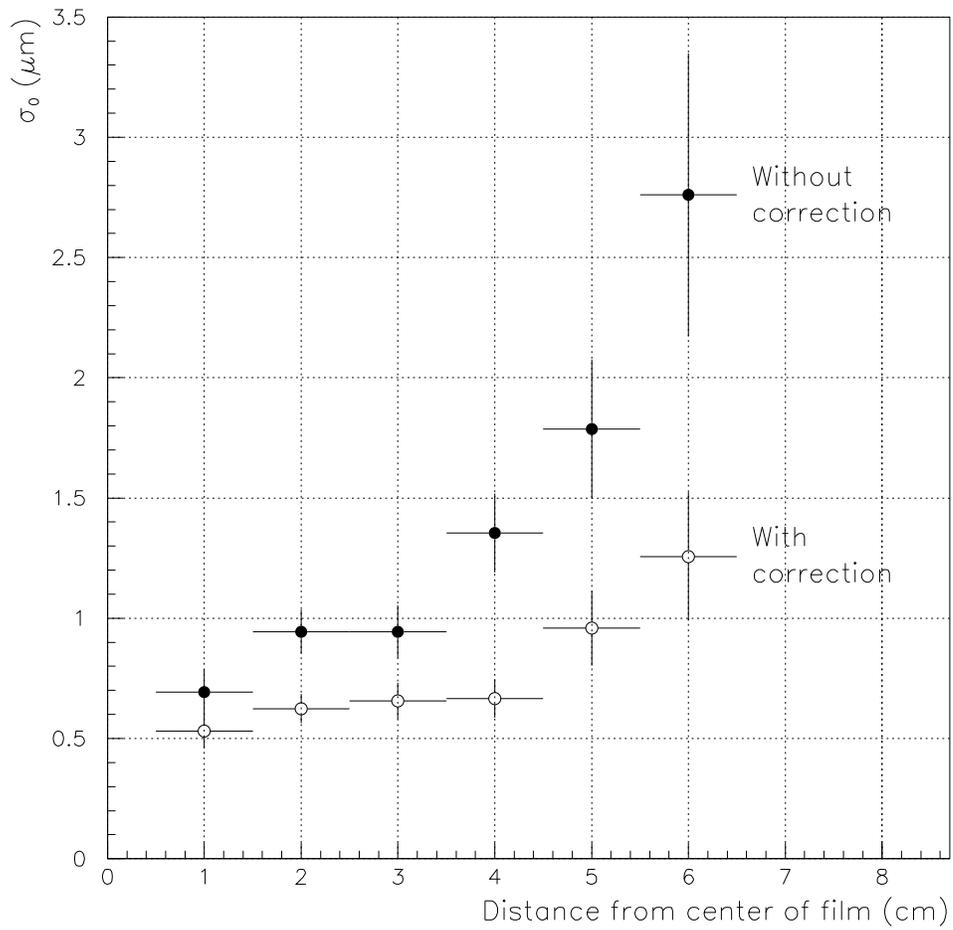}
 \end{center}
 \caption{Residuals of track positions $\sigma_0$ as a function of distance from 
the film center. Black (outlined) circles are those without (with) correction.}
 \label{fig:area_distr}
\end{figure}
\begin{figure}[htbp]
 \begin{center}
  \includegraphics[width=140mm]{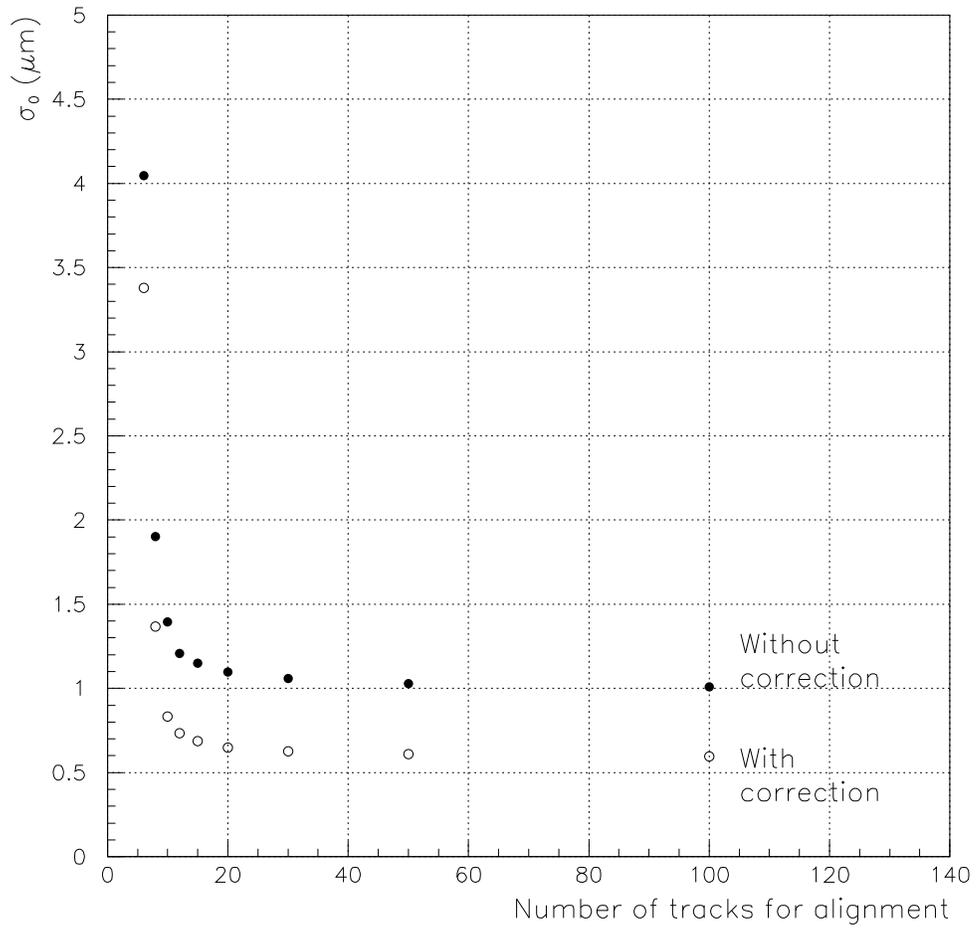}
 \end{center}
 \caption{Residuals of track positions $\sigma_0$ as a function of number of
reference tracks used for alignment. 
Black (outlined) circles are those without (with) correction.}

 \label{fig:density}
\end{figure}

Then the original, exposed, track position is recovered by using the nearest grid mark. 
Since the edge part are severe deformation, we use the nearest grid mark.
The deformation correction by using the grid marks are made for 2D vector shift.
We compare the residuals of muon positions in $x$ direction with or without this correction 
in Fig.~\ref{fig:area_distr}. Small, less than 1~\textmu m, and almost constant residuals 
are obtained in the central region. We still notice some deterioration 
of position resolution in the edge region. We think those are mostly due to 
deformation of the emulsion films in the brick package. Those should be avoided in the future 
by an improved packing procedure. Meanwhile, we confine the rest of the analysis
to the central 5 cm $\times$ 7 cm area of the emulsion films. 
Fig.~\ref{fig:density} shows the dependence of the alignment accuracy on the number 
of reference muon tracks used in the minimization procedure described above. 
We obtain 0.6~\textmu m residual by using 50 or more muon tracks but we
note that submicron alignment is achievable with
as few as 10 muon tracks. \par
\section{Momentum measurements}
\begin{figure}[htbp]
 \begin{center}
  \includegraphics[width=140mm]{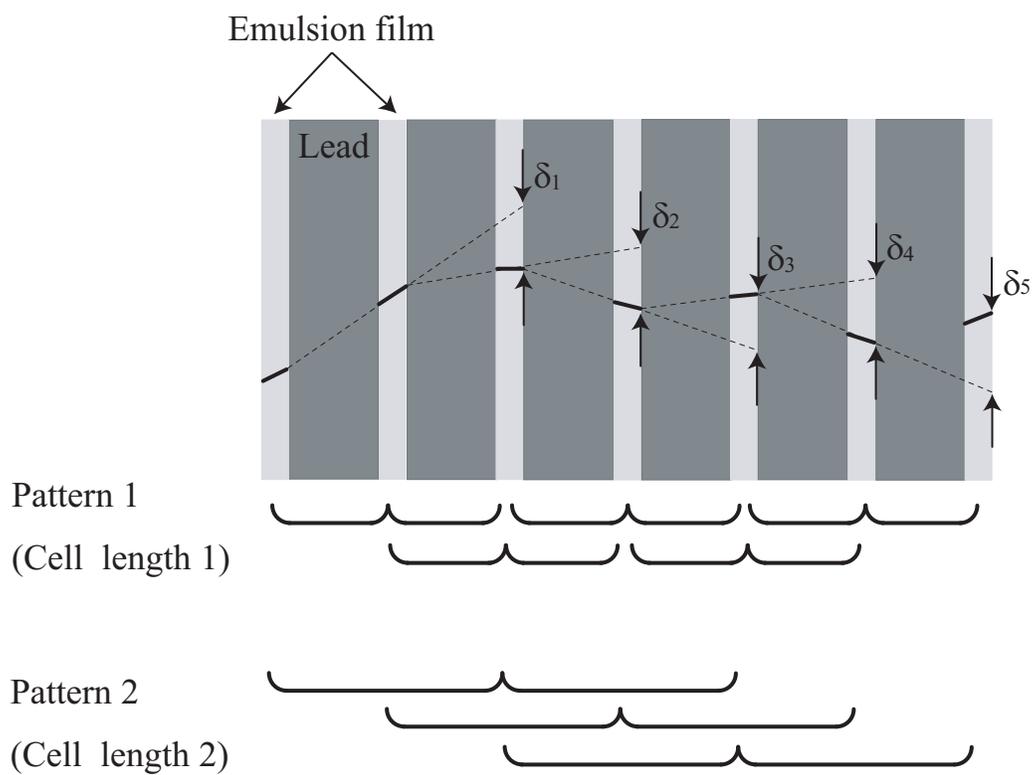}
 \end{center}
 \caption{Schematic view of the multiple Coulomb scattering measurement.}
 \label{fig:cell}
\end{figure}

In order to confirm the effectiveness of the above correction method, 
we evaluated its effect on the determination of the muon momenta of 
the different beams from their multiple Coulomb scattering in the brick.
The momentum of each muon track is estimated from position displacement by 
multiple Coulomb scattering~\cite{Coordinate}. The position displacement of each track
in one projection is expressed as
\begin{equation}
   \delta_i =  x_{i+2} - x_{i+1} - \frac{x_{i+1} - x_{i}}{z_{i+1} - z_{i}}\cdot(z_{i+2} - z_{i+1})
\end{equation}
so-called ``second difference", where the base track position at $i$th film ($i \in \{1, \ldots ,28\}$) is 
$(x_i,z_i)$.
Observable root mean square (RMS) of the second difference $\Delta_{\hbox{obs}}$ 
becomes the convolution of the signal of multiple Coulomb scattering 
$\Delta_{\hbox{sig}}$ and the measurement error $\epsilon$, and $\Delta_{\hbox{sig}}$ 
has the relationship:
\begin{equation}
    \Delta_{\hbox{sig}} = \frac{t}{2\sqrt{3}}\cdot\frac{0.0136(\mathrm{GeV})}{p\beta}\sqrt{\frac{t}{X_0}} \left\{ 1 + 0.038\ln\left(\frac{t}{X_0} \right) \right\}
\end{equation}
where $p$ and $\beta$ are the momentum and velocity of the particle, 
$t$ is the thickness of the material~\cite{PDG}. 
We use ``cell length" as unit of thickness of the material, cell length = 1 is 
a 1~mm thick lead plate plus an emulsion film, 
used to measure the second difference $\delta$. Fig.~\ref{fig:cell} shows how to obtain 
the second difference as a function of cell lengths. The second differences 
of cell length = 1 and 2 are measured by the pattern 1 and 2 respectively.\par
The measurement error $\epsilon$ of the second difference 
is composed of the position measurement accuracy 
of UTS, the alignment accuracy and the position displacement by the emulsion deformation. 
There is an approximate relation of $\epsilon = \sqrt{6}\cdot\sigma_0$ 
between the measurement error $\epsilon$ and the overall track position
measurement accuracy $\sigma_0$ at each film after alignment. 
The distribution of $\Delta_{\hbox{obs}}$ is approximately 
described by a Gaussian distribution for each muon track. The momentum of the track 
is extracted from the observed standard deviation of this distribution.
This determination can in principle be performed for different choices of the cell length. 
In our analysis, the cell length was defined as the shortest possible one for which 
the expected $\Delta_{\hbox{sig}}$ exceeds the estimated measurement error.
The application of the correction method described in section 4 allows to reduce 
the measurement error and, thus, in some cases, to reduce the cell length, which 
further contributes to improve the momentum resolution.\par
\begin{figure}[htbp]
  \begin{center}
   \includegraphics[width=140mm]{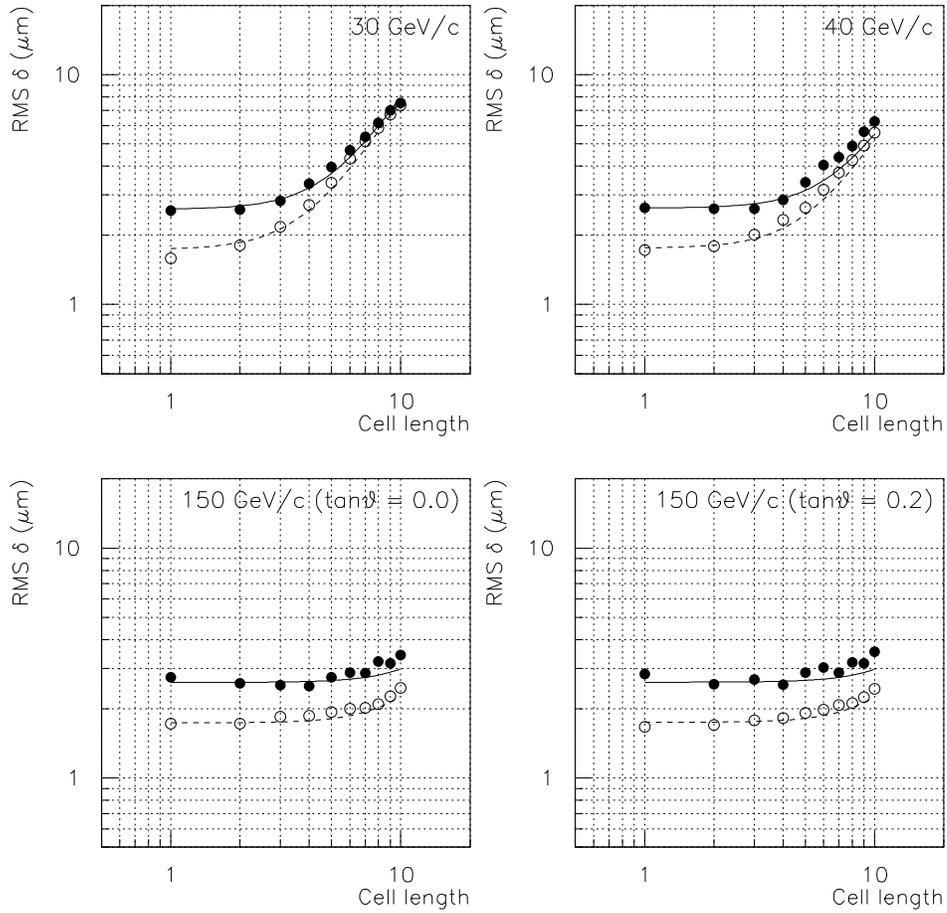}
  \end{center}
  \caption{Second difference RMS of muon tracks as a function of thickness
of the material in unit of the cell length. 
Black (outlined) circles show those without (with) correction. 
Solid (dashed) lines indicate predictions by the GEANT4 simulation
without (with) correction.}
  \label{fig:sd}
\end{figure}
\begin{figure}[htbp]
  \begin{center}
   \includegraphics[width=140mm]{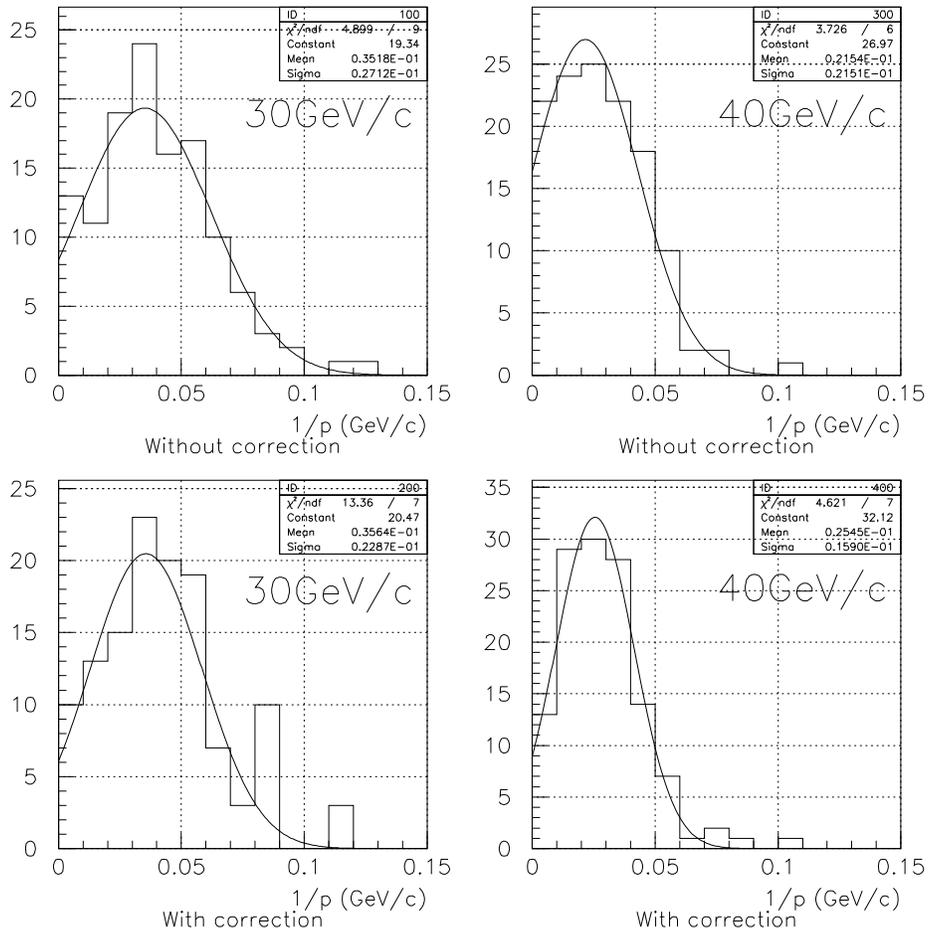}
  \end{center}
  \caption{Distributions of the inverse of momenta ($1/p$) for 30 and 40 GeV/$c$ muons. 
Solid lines represent fit results with Gaussian functions. }

  \label{fig:moment}
\end{figure}

Second differences, the position differences at the film intervals, without (with) 
correction in $x$ direction are shown in Fig.~\ref{fig:sd} for each muon beam, 
where the results obtained from a GEANT4~\cite{GEANT4} based simulation are overlaid.
The GEANT4 simulation includes residuals obtained from this analysis.
Since an angle of track is larger in $y$ projection, the momentum is estimated from the second difference in $x$ projection~\cite{Angular2}.
Total behaviors of second differences with 
respect to the cell lengths are in agreement with the GEANT4 simulation. 
Fig.~\ref{fig:moment} 
shows the second difference distributions scaled to give $1/p$ for 30 and 40 GeV/$c$ 
muons. These distributions are fitted by Gaussian functions 
with standard deviations of 77\% (64\%) and 100\% (63\%) 
for those without (with) correction of 30 and 40~GeV/$c$ muons, respectively.\par
\section{Conclusions}
New correction method of the emulsion deformation has been developed to recover the spatial resolution
of the nuclear emulsion over a large area. The method recovers the position resolution deterioration
due to the deformation mostly in the development process. The method employs a precise photomask,
careful treatment of nuclear emulsion, and alignment technique among emulsion films in order to
enhance the positioning accuracy. Position measurement accuracy of 0.6~\textmu m is obtained
over the area of 5~cm $\times$ 7~cm. The method allows to measure positions of track segments
with submicron accuracy in an ECC brick with 10 reference tracks for alignment.
It will be applied to long baseline neutrino oscillation experiments where
only a few reference tracks for alignment are expected.
It will also be useful for other applications of emulsion films as a tracking device
such as the emulsion spectrometer technique
proposed for future neutrino oscillation experiments.

\section*{Acknowledgments}
We would like to thank the colleagues of the Fundamental Particle Physics Laboratory, 
Nagoya University for their cooperation.
We gratefully acknowledge the financial supports from the Promotion and Mutual Aid Corporation
for Private Schools of Japan and the Futaba Electronics Memorial Foundation.
We thank HOYA corporation for providing the photomask used in this study. 
For the beam exposure, we acknowledge the support of the SPS staff at CERN.
We would like to thank P. Vilain for his careful reading of the manuscript.

%




\newpage

\end{document}